\documentclass[prd,twocolumn,showpacs, preprintnumbers, showkeys]{revtex4}
\usepackage{graphicx}
\begin{document}
\preprint{ANL-HEP-PR-08-34}
\title{Equilibrium criterion and effective spin models for finite temperature gauge theories}

\author{Alexander Velytsky}
\email{vel@theory.uchicago.edu}
\affiliation{Enrico Fermi Institute, University of Chicago, 5640 S. Ellis Ave., Chicago, IL 60637, USA,}
\affiliation{HEP Division and Physics Division, Argonne National Laboratory, 9700 Cass Ave., Argonne, IL 60439, USA}
\date{\today}


\begin{abstract}
Using the example of the $SU(2)$ gauge theory in $3+1$ dimensions we consider 
the construction of a $3$-dimensional effective model in terms  
of Polyakov loops. We demonstrate the application
of an equilibrium self-consistency condition to the systematic analysis of 
the contribution of
various (global $Z(2)$ symmetric) terms in the effective model action. 
We apply this analysis to the construction of a simple effective action 
with the minimum necessary number of operators. Such an action is shown to
be capable of reproducing relevant observables, e.g. the Polyakov loop 
ensemble average, within the desired accuracy.
\end{abstract}


\maketitle

\section{Introduction}
There has been renewed interest in the construction of effective models for
the finite temperature confinement-deconfinement phase transition of 
non-abelian $SU(N)$ gauge theories, see Ref.  \cite{Wozar:2007tz}  for $SU(3)$ and 
Refs. \cite{Dittmann:2003qt,Heinzl:2005xv} for $SU(2)$. The idea is to use relevant variables
to rewrite the gauge theory as an effective lower dimensional spin model with the goal of simplifying the description of phase
transition dynamics. For $SU(N)$ gauge theories the center group global 
symmetry $Z(N)$ controls the transition: it is spontaneously broken in the deconfined phase but unbroken in the confined phase.   
Therefore, the effective model can
be expressed in terms of the relevant order parameter (Polyakov loop) \cite{Polyakov:1978vu,Susskind:1979up}: 
\begin{equation}
{\cal P}_x=\prod_{n=1}^{N_t}U_{x+n\hat{t},0}, \quad L(x)=\frac1N Tr{\cal P}_x.
\end{equation}
For $SU(2)$ gauge theory the transition is second order and the universality 
argument holds. It is conjectured that for $N>2$ groups 
the integration of all degrees of freedom but the order parameter leads
to a short-range effective model \cite{Svetitsky:1982gs}.

The successful construction of such an effective model results in significant
simplification of the description of the physics. Also, in practical terms, 
it is much easier to perform
Monte Carlo simulations of a $(d-1)$-dimensional spin 
system than a $d$-dimensional gauge theory.

Using pure symmetry 
considerations
it is straightforward to identify an infinite set of possible operators 
that may enter in the effective model construction. As the next step,  
it is necessary to order these operators by their importance. 
This would allow for  a  
discriminative truncation of the effective action to a manageable minimal set of operators.
It is possible to use the strong coupling and character expansions, together 
with natural truncation criteria,  
to sort 
the hierarchy of operators of a given interaction range \cite{Billo:1996wv}.
The interplay, however, of truncating in the coupling versus the 
range of hopping terms in the action cannot 
be resolved by this approach.

In general, truncations result in an action which is capable only of an 
approximate description of the original gauge theory ensemble. Passing to the effective model defined by the truncated action may introduce non-equilibrium effects (thermalization). This is especially important since the effective theory parameters are estimated starting from the gauge theory ensemble, which may not be a representative
equilibrium ensemble of the effective action.  Therefore,  it is important to control residual non-equilibrium effects.

The most interesting aspect of the study of non-equilibrium effects, however, 
is that it provides a natural classification of the importance of various effective model operators. This is achieved by adhering to the following procedure: 
\begin{itemize}
\item From a given ensemble of 4D gauge configurations $U_{x,x_0}$ produced by standard gauge theory methods (heatbath/overrelaxation), we extract a 3D field of Polyakov loops $\mathcal{P}_x$. 
\item An effective action is chosen by specifying a particular set of 
operators.  
\item The system is allowed to evolve microcanonically along the sheet of constant energy measured according to the effective action. 
\item The process of equilibration is observed. Particularly we measure the evolution of various physical observables as they thermalize at new values. 
\end{itemize}

The microcanonical evolution is realized through demon updating \cite{Creutz:1983ra,Hasenbusch:1994ne}. 
We conduct our study over a range of lattice couplings. Since, however, 
the normalization procedure for the demon energy is unknown we decided to study only the ensemble average of the  fundamental 
Polyakov loop $\langle L\rangle$. Its evolution is monitored as the system is evolved toward equilibrium. We quantify the overall non-equilibrium effect by 
the difference between the final (equilibrium) value and a starting (gauge ensemble) value $\delta \langle L\rangle$. If one could chose an effective action capable of reproducing the dynamics of Polyakov loops exactly, then no discernible change in observalbes would be observed.
By adding or removing operators and then observing the change, we can sort 
these operators by their importance.
The truncation criteria
can be set so as to minimize the non-equilibrium effects (thermalization flow) in the 
transition to the effective model, starting from an ensemble generated
with the gauge theory Lagrangian. 
This approach is  
similar to equilibrium self-consistency ideas used in previous MCRG 
decimation studies \cite{Tomboulis:2007rn,Tomboulis:2007re,Tomboulis:2007nn}. 

Note that the use of the demon method allows for easy monitoring of observables while performing
the measuremts. This is in contrast to previous studies relying on the 
Schwinger-Dyson method 
for measurements of effective couplings. In that method "black box" measurements are performed without any control of
non-equilibrium effects. One could improve
on the SD procedure by performing the Monte-Carlo evolution with the new measured action starting from the gauge configurations, while observing the equilibration of observables. Then the analysis of effective action terms similar to the one proposed here should be used with the end goal of avoiding any equilibration changes in observalbes.

It should also be noted 
that the equilibration effect can be easily missed on small lattices: $20^3\times4$ shows almost no effect.

This approach will be
pursued in this paper in the case of the $SU(2)$ gauge theory. The same
method is also applicable to other $SU(N)$ gauge theories, and, in particular,
$SU(3)$ gauge theory.

\section{The simplest one-operator effective action}
It is natural to start building an effective model considering only the simplest possible
term. Therefore, in this section we focus on the fundamental character nearest neighbor hopping term
\begin{equation}
\hat{A}_1=\chi_{1/2}({\cal P}_x)\chi_{1/2}({\cal P}_y).\label{eq:simpleact}
\end{equation}

\begin{figure}[ht]
\includegraphics[width=\columnwidth]{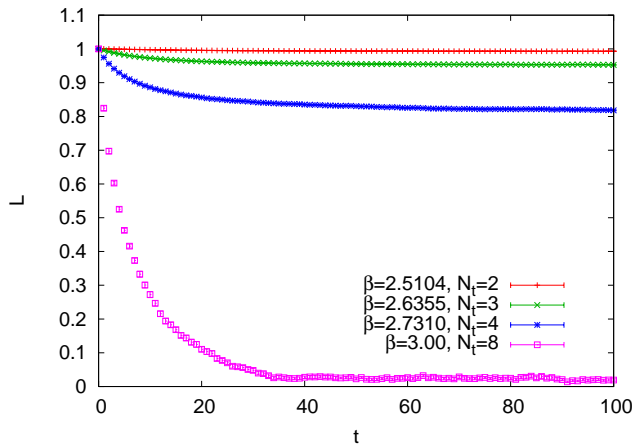}
\caption{The renormalized Polyakov loop average at $T=4T_c$ on different $40^3\times N_t$ lattices.}
\label{fig:ploop_4tc}
\end{figure}

After preparing a typical gauge configuration at gauge coupling $\beta$ we apply the demon method
assuming the effective action 
$S_1=\lambda_1\hat{A}_1$.
The demon was originally thermalized for 100 sweeps on an equivalent gauge configuration which was then discarded. In this way we remove the effect of demon thermalization. In Fig. \ref{fig:ploop_4tc} we present the evolution of the Polyakov loop average $\langle L\rangle$
at fixed physical temperature $T=4T_c$ and different lattice couplings. 
Since the renormalization of the Polyakov loop is multiplicative, in order to compare the relative flow scales at different $\beta$ we normilize the Polyakov loop to unity. One notable feature of the plot is that the Polyakov loop shows vanishingly small change as $\beta$ values become smaller. 

In all measurements presented here there are several independent simulations which allow us to extract errors using the jack-knife method.
For this simulation we take $\beta$ values to correspond to critical coupling values at some integer $N_t$ (see the last column in Tab. \ref{tab:beta_drop}), which were estimated in \cite{Velytsky:2007gj}.
This makes fixing the temperature very straightforward. 

In Tab. \ref{tab:beta_drop} we present the data for the relative change in the Polyakov loop 
$\delta\langle L\rangle/\langle L\rangle$ after 1000 sweeps evolution.
\begin{table}
\caption{\label{tab:beta_drop} The relative change of the Polyakov loop ensemble average 
$\delta L/L\equiv\delta\langle L\rangle/\langle L\rangle$ after 1000 Monte-Carlo sweeps at various temperatures and inverse lattice couplings on  $40^3\times N_t$ lattices. The starred (*) value is computed using the weak coupling 1 loop formula.}
\centering
{\small
\begin{tabular}{c|cc|cc|c}
&\multicolumn{2}{|c|}{$T=4T_c$}&\multicolumn{2}{|c|}{$T=2T_c$}&$T=T_c$\\\hline
$\beta$ &$N_t$&$\delta L/L$&$N_t$&$\delta L/L$&$N_t$\\ \hline
3.00$^*$&8&0.974(4)&&&32\\
2.7310    &4&0.191(1)&8&0.973(3)&16\\
2.6355    &3&0.0501(4)&6&0.975(3)&12\\
2.5104    &2&0.0073(1)&4&0.57(1)&8\\
2.2991    & &                                     &2&0.0153(3)&4\\
1.187348& &         & &                                     &2\\
\end{tabular}
}
\end{table}
It shows insignificant flow as one approaches the strong coupling region while the flow in the  
the weak coupling region is considerable.

In the limits of strong and weak coupling the considered action is expected
to reproduce the physics of the underlying gauge theory \cite{Ogilvie:1983ss}. 
It is possible to estimate the coupling of the spin model $\lambda_1$ using perturbative expansion.
In the strong coupling limit it is \cite{Ogilvie:1983ss}, \cite{Billo:1996wv}\footnote{Note factor 2 difference between notation used in these papers (we use Billo notation).} 
\begin{equation}
\lambda_1=(\beta/4)^{N_t}.
\end{equation}
In the weak coupling one expects
\begin{equation}
\lambda_1=\beta/(N^2N_t)=\beta/(4N_t).
\end{equation}

\begin{figure}[ht]
\includegraphics[width=\columnwidth]{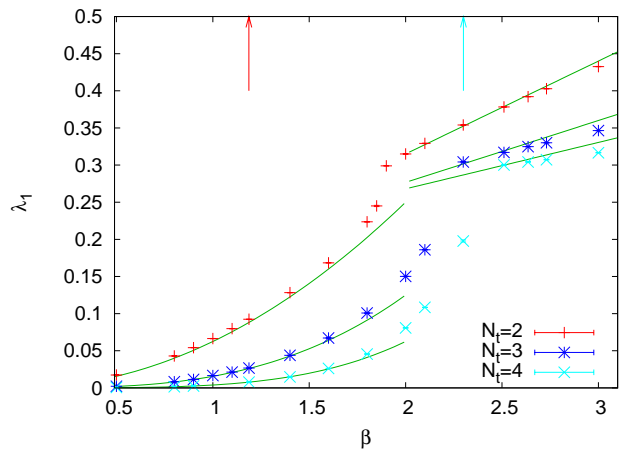}
\caption{Effective spin model coupling as function of gauge coupling.
Arrows indicate  the critical coupling values for lattices $N_t=2$ and 4.}
\label{fig:l1}
\end{figure}

This part of our study is similar to \cite{Gocksch:1984ih}. 
We measure the effective coupling using the demon method. For this we allow 
the demon to thermalize the system for 500 sweeps,
and then monitor the demon energy value for the next 10 sweeps. We present the values of coupling in Tab. \ref{tab:l1} for various $\beta$ values and 3 lattices.
\begin{table}[ht]
\caption{\label{tab:l1}Effective coupling for the simple one-operator model as a function of the gauge coupling for $N_t=2,3$ and 4 lattices. Last rows show the leading coefficients of the fits.}
\begin{center}
\begin{tabular}{c|c|c|c}
$\beta$&$N_t=2$&$N_t=3$&$N_t=4$\\\hline
0.5 &0.0174(4)&0.0022(2)&0.0006(2)\\
0.8 &0.0430(2)&0.0082(2)&0.0016(4)\\
0.9 &0.0542(2)&0.0114(4)&0.0026(2)\\
1.187348&0.0924(1)&0.0267(4)&0.0078(2)\\
1.6 &0.1684(2)&0.0672(2)&0.0262(2)\\
1.8 &0.2236(4)&0.1008(2)&0.0456(4)\\
2.0 &0.3150(2)&0.1502(2)&0.0808(2)\\
2.2991&0.3540(1)&0.3042(2)&0.1978(14)\\
2.5104&0.3783(1)&0.3170(2)&0.3001(1)\\
2.6355&0.3921(1)&0.3248(2)&0.3042(2)\\
2.7310&0.4028(1)&0.3300(4)&0.3072(2)\\ 
3.0 & 0.4326(2)&0.3464(2)&0.3166(2)\\\hline
$a_s$&0.0642(4)&0.0159(3)&0.0038(1)\\
$1/4^{N_t}$& 0.0625&0.0156&0.0039\\\hline
$a_w$&0.1109(1)&0.0599(5)&0.0334(5)\\
$1/4N_t$&0.125&0.083&0.625\\
\end{tabular}
\end{center}
\end{table}%
In Fig. \ref{fig:l1} we plot the effective coupling values. Note that the figure contains more data points than the table. 
We check the weak and strong coupling behavior against the perturbation theory results.
For the strong coupling we perform the fit to $a_s\beta^{N_t}+c$, while for the weak coupling the functional form is $a_w\beta+c'$. The results of the fits are presented in the lower part of Tab. \ref{tab:l1}.
We indicate by solid lines in Fig. \ref{fig:l1} the perturbation theory results.

We observe that at strong coupling the perturbation scaling is indeed 
achieved. This nicely corresponds to the fact that there is minimal flow in this region. For weak coupling we observe deviation from 
the perturbative result (similar observations were made in \cite{Gocksch:1984ih} and \cite{Heinzl:2005xv}). This is easy to understand since in this region 
we observed significant flow  
of the Polyakov loop during the thermalization.   
Note that smaller time extent lattices show results closer to the perturbation theory behavior in the weak coupling regime. The smallest  $N_t=2$ lattice indeed shows fairly good agreement in this region.

We conclude that the one-operator action is capable of describing the gauge theory in the strong coupling limit. In the weak coupling limit, however, 
it represents a good approximation only for small time extent lattices $N_t=1$ and 2. This result is not surprising since the gauge theory at weak coupling exhibits  non-local nature which cannot be captured by a local effective action.

\section{Analysis of general effective action}
Next we fix the coupling value to $\beta=2.7310$. This is a value from the region where, as we saw in the previous section, the simplest action exhibited a significant flow in the Polyakov loop average. On $40^4\times 4$ lattice it corresponds to fixed temperature $4T_c$. We start adding higher order terms to the simplest action 
in order to reduce the flow. It is natural to group the terms by the spatial 
range of the hopping terms.
The nearest neighbor hopping term group $y=x+\hat\mu$ is 
\begin{eqnarray}
\hat A_1=&\chi_{1/2}({\cal P}_x)\chi_{1/2}({\cal P}_y),\quad &O(L^2)\\
\hat A_2=&\chi_{1}({\cal P}_x)\chi_{1}({\cal P}_y),\quad &O(L^4)\\
\hat A_3=&\chi_{3/2}({\cal P}_x)\chi_{1/2}({\cal P}_y),\quad &O(L^4)\\
\hat A_4=&\chi_{3/2}({\cal P}_x)\chi_{3/2}({\cal P}_y),\quad &O(L^6)\\
\hat A_5=&\chi_{2}({\cal P}_x)\chi_{1}({\cal P}_y),\quad &O(L^6).
\end{eqnarray}
The potential terms group is
\begin{eqnarray}
\hat B_1=&\chi_1({\cal P}_x),\quad &O(L^2)\\
\hat B_2=&\chi_2({\cal P}_x),\quad &O(L^4)\\
\hat B_3=&\chi_3({\cal P}_x),\quad &O(L^6).
\end{eqnarray}
Here the order of operators $O(L^n)$ is specified by representing characters as  $n$th order polynomials of the fundamental representation loop $L$.
The next to nearest neighbor group is
\begin{eqnarray}
\hat C_1=&\chi_{1/2}({\cal P}_x)\chi_{1/2}({\cal P}_y),\\
\hat C_2=&\chi_{1}({\cal P}_x)\chi_{1}({\cal P}_y),\\
\hat C_3=&\chi_{3/2}({\cal P}_x)\chi_{1/2}({\cal P}_y).
\end{eqnarray}

\begin{figure}[ht]
\includegraphics[width=\columnwidth]{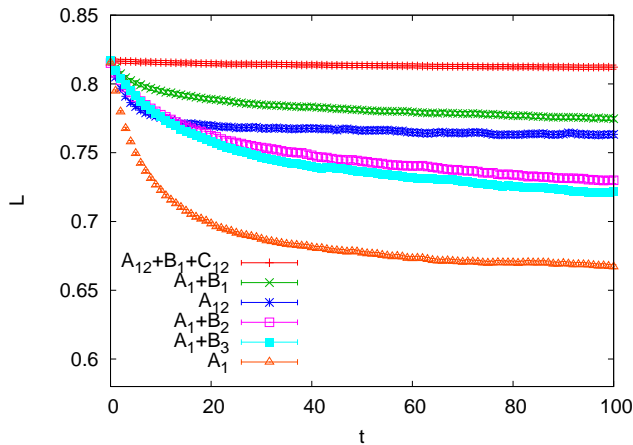}
\caption{The Polyakov loop flow for the simplest one-operator action, various two-operator actions and the $A_{12}+B_1+C_{12}$ action on a $T=4T_c$ $\beta=2.7310$ lattice.}
\label{fig:ploop_4}
\end{figure}

By analogy we introduce $\hat D_i(x,y)$ for the next to next to nearest neighbor terms $y=x+\hat\mu+\hat\nu+\hat \kappa$ and $\mu\neq\nu\neq\kappa$.
We consider various actions of the form
$S_i=\lambda_{O_i} \hat O_i$, where $\hat O$ is any of $A, B,C$ or $D$.
For simplicity we introduce the shorthand notation $O_{i...j...k}=S_1+...+S_j+...+S_k$, thus $A_{12}=\lambda_{A_1}\hat A_1+
\lambda_{A_2} \hat A_2$.
\begin{table}[ht]
\caption{\label{tab:lchange}The change of the Polyakov loop $L$ after 1000 sweeps for various combinations of terms in the effective action. The original gauge theory configuration is at $\beta=2.7310$ on a $40^4\times 4$ lattice.}
\begin{center}
\begin{tabular}{cc}
$S$           & $\delta L$\\
$A_1$       & 0.1556(1)\\
$A_{1}+B_1$  &0.047(1)\\
$A_{12}+B_1$&0.031(1)\\
$A_{13}+B_1$    &0.045(1)\\
$A_{1}+B_{12}$    &0.044(1)\\
$A_{123}+B_{12}$ &0.034(1)\\
$A_{1234}+B_{12}$&0.0307(5)\\
$A_{1235}+B_{12}$&0.0317(5)\\
$A_{123}+B_{123}$&0.0309(2)\\
$A_{12345}+B_{123}$&0.0286(4)\\ \hline
$A_{1}+B_1+C_{1}$    & 0.0142(4)\\
$A_{12}+B_1+C_{1}$  &0.0102(3)\\
$A_{12}+B_1+C_{12}$&0.0065(2)\\
$A_{12345}+B_{123}+C_{123}$&0.0041(1)\\ \hline
$A_1+B_1+C_1+D_1$&0.0102(1)\\
$A_{12}+B_1+C_{12}+D_{1}$&0.0048(1)\\
$A_{12}+B_1+C_{12}+D_{12}$&0.0042(1)\\
$A_{12345}+B_{123}+C_{123}+D_{12}$&0.0049(1)
\end{tabular}
\end{center}
\end{table}%

It is impractical to consider all combinations of the indicated operators in the action. 
Instead we consider only several combinations which reveal the relative importance of the terms.
We list in Tab. \ref{tab:lchange} the change in the Polaykov loop average for different combinations of the operators.
If one sets a goal to realistically reproduce the value of the Polyakov loop within 1\% precision then the 
$A_{12}+B_1+C_{12}$ effective action is the minimal action which is suitable.
We also demonstrate the relative contribution of various  terms in Fig. \ref{fig:ploop_4}, where different two-operator effective actions are compared to the simplest one-operator action
and the $A_{12}+B_1+C_{12}$ action. We see that the most significant improvement comes from the $B_1$ term, while the next significant term is $A_2$.

Next we measure the couplings of the $A_{12}+B_1+C_{12}$ effective model 
at different $\beta$ values and corresponding temperatures
on a $40^3\times 4$ lattice, see Tab. \ref{tab:terms_b}. Note that among 
the couplings measured in the confined phase only 
potential terms and the fundamental hopping term have significant values. 
Also, for comparison, we present in the table the couplings for the effective model with all 13 operators (computed at $\beta=2.7310$). We use the same statistics for this measurement. It is obvious that the errors are much larger for the full action. We also note little change in the value of the couplings present in both the full and reduced $A_{12}+B_1+C_{12}$ effective models. It is interesting that some of the operators present in the full action but 
not present in the reduced action have couplings which are significant, 
e.g. $\lambda_{A_3},\lambda_{B_2},\lambda_{D_1}$. However this terms effect on the Polyakov loop average turns out to be small.

\begin{table}[ht]
\caption{The couplings of the $A_{12}+B_1+C_{12}$ effective  action at various $\beta$ and of the full
13-operator action at $\beta=2.7310$ on a 
$40^3\times 4$ lattice. 
Demon measurements over 10 sweeps after 100 sweeps of thermalization (10 runs).}
\label{tab:terms_b}
\centering
\begin{tabular}{c|cccc|c}
$\beta$&1.187348&2.2991&2.5104 & 2.7310& 2.7310\\\hline
$\lambda_{A_1}$&0.0064(4)&0.1252(3)& 0.1444(7)&0.1666(6)&0.1692(55)\\
$\lambda_{B_1}$&-0.0436(2)&-0.0578(2)&-0.0679(3)&-0.0745(3)&-0.0742(14)\\
$\lambda_{A_2}$&-0.0008(4)&-0.0049(4)&-0.0120(7)&-0.0110(5)&-0.0078(31)\\
$\lambda_{C_1}$&-0.0002(3)&0.0004(4)& 0.0409(4)&0.0389(3)&0.0225(30)\\
$\lambda_{C_2}$&-0.0002(4)&-0.0020(1)&-0.0069(4)&-0.00057(4)&-0.0041(19)\\
$\lambda_{A_3}$&&&&&0.0100(29)\\
$\lambda_{B_2}$&&&&&-0.01377(14)\\
$\lambda_{A_4}$&&&&&0.0013(29)\\
$\lambda_{A_5}$&&&&&-0.0025(14)\\
$\lambda_{B_3}$&&&&&-0.0037(6)\\
$\lambda_{C_3}$&&&&&0.0008(13)\\
$\lambda_{D_1}$&&&&&0.0199(29)\\
$\lambda_{D_2}$&&&&&-0.0007(21)\\
\end{tabular}
\end{table}

\begin{figure}[ht]
\includegraphics[width=\columnwidth]{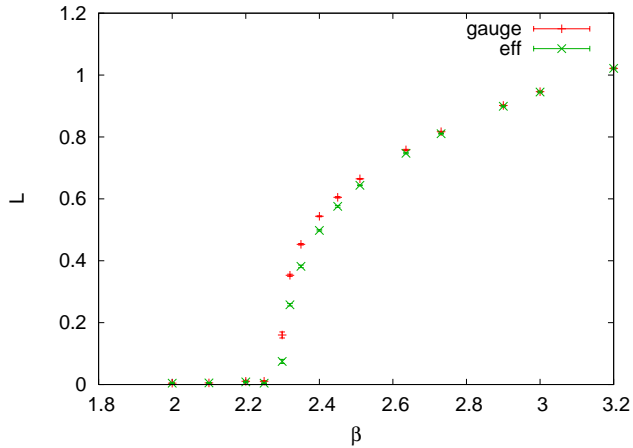}
\caption{The Polyakov loop average computed from the $A_{12}+B_1+C_{12}$ effective model and the gauge theory
on a $40^3\times 4$ lattice.}
\label{fig:ploopb}
\end{figure}

In Fig. \ref{fig:ploopb} we check the action $A_{12}+B_1+C_{12}$ over a wide range of couplings. We observe that the effective action is tracing the value of the Polyakov loop average computed from the gauge theory very closely. The agreement is very good in the weak coupling regime. Near the transition region 
there is small discrepancy in the values; however, the location of the 
phase transition is accurate.

\section{Summary}
We presented a simple empirical method of selecting relevant terms in the construction of an effective spin action capable of reproducing a single observable (the Polyakov loop average). By choosing operators for the effective action and measuring their couplings using the demon method, we were able to construct and test various effective action models.
The simplest one-operator effective action was analyzed with respect to 
the size of non-equilibrium effects of thermalization of the expectation value of the Polyakov loop. It was also compared to perturbation theory results. We find that the strong coupling regime, where the perturbation theory results correspond to the measured 
observable values, is the regime of minimal flow. The weak coupling regime exhibits serious non-equilibrium effects, which indicates that in this region the one-operator action is not appropriate for the description of Polyakov loop configurations. 
More general effective actions with up to 13 operators were considered and analyzed with respect to Polyakov loop flow. Setting as a criterion to reproduce   
the expectation value of the Polyakov loop with accuracy under 1\%, we found that it is enough to limit the effective action to five operators $A_{12}+B_1+C_{12}$. We showed that this action is capable of correctly reproducing the value of the Polyakov loop average over a wide range of lattice couplings.

We should note that it is also possible to consider different observables. 
This would, in general, require a construction of a new action. It is obvious that for long distance observables, such as Polyakov loop correlators, one would need a more complex effective action model.

\section*{Acknowledgments}
The author would like to thank T. Tomboulis for insightful discussions. The computations were performed on the cluster maintained by 
Academic Technology Services (UCLA).
This work is supported in part by the U.S. Department of Energy, Division of High Energy Physics and Office of Nuclear Physics, under Contract DE-AC02-06CH11357, and in part under a Joint Theory Institute grant.

\bibliographystyle{apsrev}
\bibliography{avref} 

\begin{thebibliography}{15}
\expandafter\ifx\csname natexlab\endcsname\relax\def\natexlab#1{#1}\fi
\expandafter\ifx\csname bibnamefont\endcsname\relax
  \def\bibnamefont#1{#1}\fi
\expandafter\ifx\csname bibfnamefont\endcsname\relax
  \def\bibfnamefont#1{#1}\fi
\expandafter\ifx\csname citenamefont\endcsname\relax
  \def\citenamefont#1{#1}\fi
\expandafter\ifx\csname url\endcsname\relax
  \def\url#1{\texttt{#1}}\fi
\expandafter\ifx\csname urlprefix\endcsname\relax\def\urlprefix{URL }\fi
\providecommand{\bibinfo}[2]{#2}
\providecommand{\eprint}[2][]{\url{#2}}

\bibitem[{\citenamefont{Wozar et~al.}(0400)\citenamefont{Wozar, Kaestner, Wipf,
  and Heinzl}}]{Wozar:2007tz}
\bibinfo{author}{\bibfnamefont{C.}~\bibnamefont{Wozar}},
  \bibinfo{author}{\bibfnamefont{T.}~\bibnamefont{Kaestner}},
  \bibinfo{author}{\bibfnamefont{A.}~\bibnamefont{Wipf}}, \bibnamefont{and}
  \bibinfo{author}{\bibfnamefont{T.}~\bibnamefont{Heinzl}}
  (\bibinfo{year}{0400}), \eprint{arXiv:0704.2570 [hep-lat]}.

\bibitem[{\citenamefont{Dittmann et~al.}(2004)\citenamefont{Dittmann, Heinzl,
  and Wipf}}]{Dittmann:2003qt}
\bibinfo{author}{\bibfnamefont{L.}~\bibnamefont{Dittmann}},
  \bibinfo{author}{\bibfnamefont{T.}~\bibnamefont{Heinzl}}, \bibnamefont{and}
  \bibinfo{author}{\bibfnamefont{A.}~\bibnamefont{Wipf}},
  \bibinfo{journal}{JHEP} \textbf{\bibinfo{volume}{06}}, \bibinfo{pages}{005}
  (\bibinfo{year}{2004}), \eprint{hep-lat/0306032}.

\bibitem[{\citenamefont{Heinzl et~al.}(2005)\citenamefont{Heinzl, Kaestner, and
  Wipf}}]{Heinzl:2005xv}
\bibinfo{author}{\bibfnamefont{T.}~\bibnamefont{Heinzl}},
  \bibinfo{author}{\bibfnamefont{T.}~\bibnamefont{Kaestner}}, \bibnamefont{and}
  \bibinfo{author}{\bibfnamefont{A.}~\bibnamefont{Wipf}},
  \bibinfo{journal}{Phys. Rev.} \textbf{\bibinfo{volume}{D72}},
  \bibinfo{pages}{065005} (\bibinfo{year}{2005}), \eprint{hep-lat/0502013}.

\bibitem[{\citenamefont{Polyakov}(1978)}]{Polyakov:1978vu}
\bibinfo{author}{\bibfnamefont{A.~M.} \bibnamefont{Polyakov}},
  \bibinfo{journal}{Phys. Lett.} \textbf{\bibinfo{volume}{B72}},
  \bibinfo{pages}{477} (\bibinfo{year}{1978}).

\bibitem[{\citenamefont{Susskind}(1979)}]{Susskind:1979up}
\bibinfo{author}{\bibfnamefont{L.}~\bibnamefont{Susskind}},
  \bibinfo{journal}{Phys. Rev.} \textbf{\bibinfo{volume}{D20}},
  \bibinfo{pages}{2610} (\bibinfo{year}{1979}).

\bibitem[{\citenamefont{Svetitsky and Yaffe}(1982)}]{Svetitsky:1982gs}
\bibinfo{author}{\bibfnamefont{B.}~\bibnamefont{Svetitsky}} \bibnamefont{and}
  \bibinfo{author}{\bibfnamefont{L.~G.} \bibnamefont{Yaffe}},
  \bibinfo{journal}{Nucl. Phys.} \textbf{\bibinfo{volume}{B210}},
  \bibinfo{pages}{423} (\bibinfo{year}{1982}).

\bibitem[{\citenamefont{Billo et~al.}(1996)\citenamefont{Billo, Caselle,
  D'Adda, and Panzeri}}]{Billo:1996wv}
\bibinfo{author}{\bibfnamefont{M.}~\bibnamefont{Billo}},
  \bibinfo{author}{\bibfnamefont{M.}~\bibnamefont{Caselle}},
  \bibinfo{author}{\bibfnamefont{A.}~\bibnamefont{D'Adda}}, \bibnamefont{and}
  \bibinfo{author}{\bibfnamefont{S.}~\bibnamefont{Panzeri}},
  \bibinfo{journal}{Nucl. Phys.} \textbf{\bibinfo{volume}{B472}},
  \bibinfo{pages}{163} (\bibinfo{year}{1996}), \eprint{hep-lat/9601020}.

\bibitem[{\citenamefont{Creutz}(1983)}]{Creutz:1983ra}
\bibinfo{author}{\bibfnamefont{M.}~\bibnamefont{Creutz}},
  \bibinfo{journal}{Phys. Rev. Lett.} \textbf{\bibinfo{volume}{50}},
  \bibinfo{pages}{1411} (\bibinfo{year}{1983}).

\bibitem[{\citenamefont{Hasenbusch et~al.}(1994)\citenamefont{Hasenbusch, Pinn,
  and Wieczerkowski}}]{Hasenbusch:1994ne}
\bibinfo{author}{\bibfnamefont{M.}~\bibnamefont{Hasenbusch}},
  \bibinfo{author}{\bibfnamefont{K.}~\bibnamefont{Pinn}}, \bibnamefont{and}
  \bibinfo{author}{\bibfnamefont{C.}~\bibnamefont{Wieczerkowski}},
  \bibinfo{journal}{Phys. Lett.} \textbf{\bibinfo{volume}{B338}},
  \bibinfo{pages}{308} (\bibinfo{year}{1994}), \eprint{hep-lat/9406019}.

\bibitem[{\citenamefont{Tomboulis and
  Velytsky}(2007{\natexlab{a}})}]{Tomboulis:2007rn}
\bibinfo{author}{\bibfnamefont{E.~T.} \bibnamefont{Tomboulis}}
  \bibnamefont{and} \bibinfo{author}{\bibfnamefont{A.}~\bibnamefont{Velytsky}},
  \bibinfo{journal}{Phys. Rev.} \textbf{\bibinfo{volume}{D75}},
  \bibinfo{pages}{076002} (\bibinfo{year}{2007}{\natexlab{a}}),
  \eprint{hep-lat/0702015}.

\bibitem[{\citenamefont{Tomboulis and
  Velytsky}(2007{\natexlab{b}})}]{Tomboulis:2007re}
\bibinfo{author}{\bibfnamefont{E.~T.} \bibnamefont{Tomboulis}}
  \bibnamefont{and} \bibinfo{author}{\bibfnamefont{A.}~\bibnamefont{Velytsky}},
  \bibinfo{journal}{Phys. Rev. Lett} \textbf{\bibinfo{volume}{98}},
  \bibinfo{pages}{181601} (\bibinfo{year}{2007}{\natexlab{b}}),
  \eprint{hep-lat/0702027}.

\bibitem[{\citenamefont{Tomboulis and Velytsky}(2008)}]{Tomboulis:2007nn}
\bibinfo{author}{\bibfnamefont{E.~T.} \bibnamefont{Tomboulis}}
  \bibnamefont{and} \bibinfo{author}{\bibfnamefont{A.}~\bibnamefont{Velytsky}},
  \bibinfo{journal}{Int. J. Mod. Phys.} \textbf{\bibinfo{volume}{A23}},
  \bibinfo{pages}{803} (\bibinfo{year}{2008}), \eprint{0705.0383}.

\bibitem[{\citenamefont{Velytsky}(2007)}]{Velytsky:2007gj}
\bibinfo{author}{\bibfnamefont{A.}~\bibnamefont{Velytsky}}
  (\bibinfo{year}{2007}), \eprint{arXiv:0711.0748 [hep-lat]}.

\bibitem[{\citenamefont{Ogilvie}(1984)}]{Ogilvie:1983ss}
\bibinfo{author}{\bibfnamefont{M.}~\bibnamefont{Ogilvie}},
  \bibinfo{journal}{Phys. Rev. Lett.} \textbf{\bibinfo{volume}{52}},
  \bibinfo{pages}{1369} (\bibinfo{year}{1984}).

\bibitem[{\citenamefont{Gocksch and Ogilvie}(1985)}]{Gocksch:1984ih}
\bibinfo{author}{\bibfnamefont{A.}~\bibnamefont{Gocksch}} \bibnamefont{and}
  \bibinfo{author}{\bibfnamefont{M.}~\bibnamefont{Ogilvie}},
  \bibinfo{journal}{Phys. Rev. Lett.} \textbf{\bibinfo{volume}{54}},
  \bibinfo{pages}{1772} (\bibinfo{year}{1985}).

\end{thebibliography}

\end{document}